\documentclass[useAMS,usenatbib]{mn2e}
\usepackage{graphicx}
\usepackage{lscape}
\usepackage{longtable}
\usepackage{txfonts}
\title[20--100~keV properties of CVs detected in the {\em{INTEGRAL}}/IBIS Survey]{20--100~keV properties of cataclysmic variables detected in the {\em INTEGRAL}/IBIS Survey}
\author[E.J. Barlow et al. ]{E.J. Barlow,$^{1}${\thanks{E-mail:ejb@astro.soton.ac.uk}}
C. Knigge,$^{1}$ A.J. Bird,$^{1}$ A.J Dean,$^{1}$ D.J. Clark,$^{1}$ A.B. Hill,$^{1}$ 
\newauthor 
M. Molina$^{1}$ and V. Sguera$^{1}$ \\
$^{1}$School of Physics and Astronomy, University of Southampton, Highfield, Southampton SO17 1BJ, United Kingdom\
}
\begin{document}

\date{Accepted . Received ; in original form }

\pagerange{\pageref{firstpage}--\pageref{lastpage}} \pubyear{2006}

\maketitle

\label{firstpage}

\begin{abstract}
Analysis of {\it INTEGRAL}/IBIS survey observations has revealed that the rare intermediate polar and asynchronous polar cataclysmic variables are consistently found to emit in the 20--100~keV energy band, whereas synchronous polars  and the common non-magnetic CVs rarely do so.   From the correlation of a candidate {\it INTEGRAL}/IBIS survey source list with a CV catalogue, 15 CV detections by IBIS have been established including a new {\it INTEGRAL} source IGR~J06253+7334.  The properties of these sources and 4 additional CV candidates are discussed in the context of their 20--100~keV emission characteristics and we conclude that the {\it INTEGRAL} mission is an important tool in the detection of new magnetic CV systems.  Furthermore, analysis of the time-averaged spectra of CVs detected by {\it INTEGRAL} indicate that although there is little difference between the spectral slopes of the different sub-types, intermediate polars may be considerably more luminous than polars in the soft gamma-ray regime.  We also present the detection of an unusual high-energy burst from V1223~Sgr discovered by inspection of the IBIS light-curve.  Additionally, we have compared the IBIS and optical AAVSO light-curves of SS Cyg and extracted IBIS spectra during single periods of optical outburst and quiescence.  We find that the 20--100~keV flux is an order of magnitude greater during optical quiescence.  This is in agreement with previous studies which show that the hard X-ray component of SS Cyg is suppressed during high accretion states.  

%The second {\it INTEGRAL}/IBIS survey contains 209 sources in the 20-100 keV energy range.  The majority of these sources are Galactic X-ray binaries, 
%but also included are a small sample of 8 cataclysmic variables. By correlating an updated IBIS candidate source list with a catalogue of 
%known CVs, at least 7 further {\it INTEGRAL} CV detections have been established.  The properties of these CVs are discussed in the context of their 20-100 keV 
%emission characteristics.  It is found that the majority of the CV systems detected in the soft gamma-ray band have a magnetic white dwarf; more 
%specifically intermediate polars and asynchronous polars, both of which are very rare classes of object.  We also present the detection of a possible high energy burst by the intermediate polar V1223 Sgr and evidence for the 20-100~keV flux of the dwarf nova SS Cyg decreasing during optical outburst.  INTEGRAL is an important mission for the discovery of intermediate polars. new IGR

\end{abstract}

\begin{keywords}
gamma-rays:observations, surveys, X-rays:binaries, cataclysmic variables
\end{keywords}

\section{Introduction}

Cataclysmic variables (CVs) are close binary systems consisting of a late-type star transfering material onto a white dwarf. The magnetic cataclysmic variables (mCVs) are a small sub-set of the total number of catalogued CV systems \citep[$<$10\%;][]{Downes01} and fall into two categories: polars (or AM Her type after the prototype system) and intermediate polars (IPs or DQ Her type).  Polars (named thus due to strong polarisation of their optical flux) contain a white dwarf (WD) that possesses a magnetic field sufficiently strong to synchronise the spin of the WD with the orbital period of the binary \citep[P$_{orb}\simeq$P$_{spin}$; for a review of polar characteristics see][]{Cropper90}.   However, there is a very small sample of polars ($\leq$5 objects) where the WD spin period is out of synchronisation with the orbital period by a few percent, thus called asynchronous polars.  For IPs, the lack of detectable optical polarisation implies a less strong WD magnetic field, not powerful enough to enforce synchronisation between the spin and binary orbits, and typically P$_{spin}\sim0.1P_{orb}$ \citep[for an review of IPs see][]{Patterson94}.  IP nature is confirmed by the detection of coherent variability associated with the rotation of the WD.  X-ray modulations have been observed at the orbital period, spin period or a beat between the two (see, e.g.~observations of the discless IP V2400 Oph by Buckley et al. 1995).

 The WD magnetospheric radius in polar systems is large enough to prevent the formation of a disk and the infalling material is `threaded' onto the magnetic field lines creating one or two accretion regions depending on the inclination of the magnetic field with respect to the binary plane.   In IP systems, the WD magnetic field is generally not strong enough to disrupt disc formation entirely and simply truncates the inner disc, resulting in an accretion flow that is channelled down towards the magnetic poles and onto the WD surface.  In a simple model of a column of gas impacting the atmosphere of the WD, a shock will form and hard X-ray/soft gamma-ray emission will result from thermal bremsstrahlung cooling by free electrons in the hot post-shock region with kT$\sim$10s of keV. Softer X-ray (and EUV, $<$1~keV) emission is produced from the absorption and reprocessing of these photons in the WD photosphere.  Reflection of the bremsstrahlung photons at the WD surface also contributes to the hard X-ray spectrum (van Teeseling, Kaastra \& Heise 1996).  As a result, both polars and IPs should be expected to emit a hard spectrum.  In fact, in polars, the ratio of soft to hard X-ray leads to an observed excess of soft X-rays \citep{Lamb85}.  \citet{Chanmugam91} report that X-ray (2--10~keV) luminosity for IPs is greater than polars by a factor of $\sim10$.  To account for this problem, it is proposed that the high magnetic field in polars induces a `blobby' flow, since the magnetic pressure increases faster than the accretion material can adjust as it accelerates along the field lines \citep{Warner95}.  Such blobs can bypass the shock and fall straight onto the WD surface \citep{Kuijpers82}.  Alternatively, for high magnetic fields and low \.M, cyclotron cooling from fast electrons spiralling around the magnetic field lines could dominate the bremsstrahlung processes \citep{Lamb79}.  

Dwarf novae (DN) are non-magnetic CVs that are observed to undergo a temporary increase in brightness due to a thermal-viscous instability in the accretion disc \citep{Osaki74}.  Hard X-ray/soft gamma-ray emission from DNe is generally thought to originate from the boundary layer (BL) between the accretion disk and the WD surface and is related to the mass accretion rate of the disk, \.M$_{acc}$ \citep{Pringle79}.  This BL releases approximately half of the accretion energy as material is decelerated onto the WD surface.   \citet{Narayan93} find that at high accretion rates, the layer is optically thick with a temperature that decreases with decreasing \.M$_{acc}$.  Below a critical rate, \.M$_{crit}\sim$10$^{16}$~g~s$^{-1}$, however, the BL becomes optically thin and its width and temperature increases dramatically.  Hence, during DN outbursts when there is an increase in \.M$_{acc}$, the hard X-ray emission will likely be suppressed {\citep{Pringle79}.  This has indeed been observed in the bright DN SS Cyg with {\it RXTE/PCA} in the 3--12~keV band \citep{Wheatley03} and in the 20--100~keV {\it INTEGRAL} data presented here.  During the periods where \.M$_{acc}<$\.M$_{crit}$, the expanded BL can also form a hot (T$\sim10^{8}$K) corona, part of which may remain in the high accretion state, providing a source of residual high energy emission during outbursts \citep{Mcgowan04}.

%CVs are very well studied in the optical, but are not particularly well known as high-energy emitters. Nethertheless, it has long been noted with observatories such as {\it Einstein} and {\it EXOSAT} that CVs, especially the magnetic systems, can be strong X-ray emitters
%X-ray emission from CVs has been observed with several missions such as {\it Einstein} and {\it EXOSAT} \citep[for reviews see][]{Cordova95, Kuulkers06}.  
The first detections of hard X-ray emission from CVs occured in the late 1970s, most notably the {\it Astronomische Nederlandse Satelliet (ANS)} detection of SS Cyg in the 1--7~keV energy band \citep{Heise78} and the detection of the interesting IP, GK Per, by {\it Ariel V} \citep[2--18~keV energy band;][]{King79}.  Furthermore, X-ray emission from CVs has also been observed with several missions such as {\it Einstein} and {\it EXOSAT} \citep[for reviews see][]{Cordova95, Kuulkers06}. More recently, prior to {\it INTEGRAL}, there have been two missions that have detected CVs above 20~keV.  \citet{Suleimanov05} present the 3--100~keV spectra of 14 IPs obtained with {\it RXTE} and \citet{deMart04} describes the spectral and temporal properties of 4 IPs based on simultaneous soft and hard X-ray (0.1--90~keV) observations with {\it BeppoSAX}. This paper will describe the correlation procedure performed between an updated {\it {INTEGRAL}}/IBIS candidate source list and a catalogue of CVs \citep{Downes01}.  We find that at least 15 CVs are detected by IBIS in the 20--100~keV energy band, the vast majority of which are magnetic CVs.

%\begin{enumerate}
 % \item non-magnetic WD, e.g. dwarf novae
 % \item polars - strong magnetic field, no accretion disc, binary and pulsation periods synchronous to within $\sim$per cent.
 % \item intermediate polars - magnetic field strong enough to truncate inner accretion disc but not strong enough to synchronise the orbital and pulsation periods 
 % \end{enumerate}
\section[]{The {\em INTEGRAL}/IBIS survey}
The {\it INTEGRAL} satellite has been in orbit since October 2002 and to date  has completed over 400 orbits ($\sim$3 day revolutions) \citep{Winkler}.   The {\it INTEGRAL}/IBIS survey forms one of the main objectives of the {\it INTEGRAL} mission by exploiting the unprecedented imaging capability of the IBIS/ISGRI instrument in the soft gamma-ray regime \citep{Ubertini}.  The role of the survey is to expand our current knowledge of the sky in the 20--100~keV energy range, by documenting the number, distribution and behaviour of the different types of high energy emitters detected in images created from guaranteed Core Programme observations.  The IBIS survey data set consists of dedicated observations along the Galactic plane and around the Galactic centre.  Additionally, a combination of pointed and deep exposure observations taken in different parts of the sky can be added to this data set once available, eventually resulting in an all-sky survey.  The first IBIS catalogue \citep[][]{Bird04} consisted of $\sim$2500 separate observations (or science windows, ScW), while the second catalogue \citep[][]{Bird06} uses $\sim$6000.  The second catalogue was published in January 2006, reaching a flux sensitivity of $\sim$1~mCrab in the 20--100~keV energy band and listing 209 sources, of which 8 were CVs.  

This paper uses the most recent IBIS survey mosaics produced from over 18000 individual ScW, spanning the three years {\it INTEGRAL} has been in orbit (revs 46--360 approx) and processed using \textsc{OSA~v.5.1} \citep{Goldwurm03}. Staring and PV data have not been included and the noisy ScW have been filtered out based on the rms of the individual image.  These new survey maps are a considerable improvement on the maps used in the previous catalogue as a consequence of improved software (second catalogue maps were processed using \textsc{OSA~v.4.1}) and significantly increased exposure particularly away from the Galactic plane.

\section[]{Detections of CVs from IBIS survey data}

\subsection[]{Correlation with Downes catalogue}
The Catalog and Atlas of Catalysmic Variables was originally published by \citet{Downes93}, but has since been superceded by an electronically available edition  \citep{Downes01}.  As of February 2006, the catalogue contained 1830 sources, listed with their location, limited spectral informaton and subtype (including those of an unconfirmed nature, globular cluster sources and non-CVs).  Less than 10\% of the total number of CVs catalogued are classed as magnetic systems (included in the catalogue as either DQ Her or AM Her type variability) and less than 2\% are IPs ($\sim$50 sources). 

The main correlation was performed between the Downes catalogue and a preliminary IBIS excess list extracted \citep[using method described in][]{Bird06} from the most recent survey mosaic in the 20--40~keV energy band and including all excesses detected at $>$4$\sigma$.  For the purposes of the correlation exercise, a variable search radius was defined around the IBIS coordinates with a maximum value of 30$\arcmin$.  If a Downes catalogued source was found within this search radius it was flagged as a possible match.  Additional correlations were performed between the Downes catalogue and four `fake' IBIS catalogues in order to estimate the number of expected false matches, following a similar method to \citet{Stephen06}.  The fake IBIS catalogues were produced by:
\begin{enumerate}
  \item transposing the IBIS coordinates by one degree in Galactic longitude (fake 1); 
  \item mirroring IBIS coordinates in Galactic longitude (fake 2), Galactic latitude (fake 3) and both longitude and latitude (fake 4).
  \end{enumerate}

\begin{figure}
%\begin{center}
\includegraphics[totalheight=9.0cm, angle=-90]{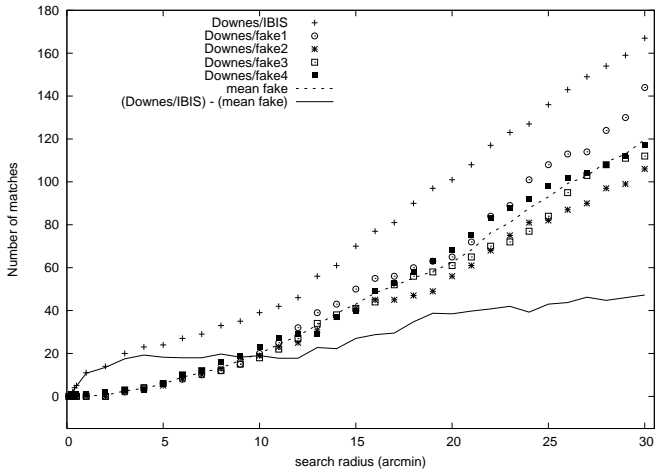}
\caption{\label{correlation} {The number of matches as a function of search radius between the Downes catalogue and the IBIS survey excess list (crosses). Also plotted are the results of correlating the Downes catalogue with 4 fake IBIS catalogues (see text for details) and the mean number of matches from the fake catalogue correlations (dotted line).  The solid line is the number of Downes/IBIS matches minus the mean of the fake matches.}}
%\end{center}
\end{figure}
 
\begin{table*}
\centering
 \begin{minipage}{140mm}
  \caption{\label{matches} {Results of Downes--IBIS correlation.}}
  \begin{tabular}{@{}l l c l c c c c@{}}
  \hline & Downes source & type & IBIS match & type & offset ($\arcmin$)\footnote{distance between the Downes catalogue coordinates and the IBIS survey coordinates} & $\sigma$\footnote{significance in the 20--40~keV IBIS survey mosaic} &  correlation result \\
 \hline
1	&	NGC 6624 CV2	&	CV	&	4U~1820-303	&	LMXB	&	0.270	&	284.7	&	FALSE	\\
2	&	V709 Cas	&	IP	&	V709 Cas	&	IP	&	0.280	&	28.2	&	TRUE	\\
3	&	NGC 6712 10261	&	CV?	&	4U~1850-087	&	LMXB	&	0.300	&	30.6	&	FALSE	\\
4	&	NGC 6652 C	&	CV?/LMXB?	&	RX~J1832-330	&	LMXB	&	0.396	&	70.2	&	FALSE	\\
5	&	IGR~J21335-5105	&	IP	&	IGR~J21335-5105	&	IP	&	0.445	&	15.1	&	TRUE	\\
6	&	IGR~J17303-6016	&	IP	&	IGR~J17303-6016	&	IP	&	0.572	&	14.5	&	TRUE	\\
7	&	V1223 Sgr	&	IP	&	V1223 Sgr	&	IP	&	0.614	&	38.3	&	TRUE	\\
8	&	CG X-1	&	P?/ULX?	&	Circinus galaxy	&	AGN	&	0.623	&	76.3	&	FALSE	\\
9	&	SS Cyg	&	DN	&	1H~2140+433	&	DN	&	0.790	&	17.7	&	TRUE	\\
10	&	V1432 Aql	&	P	&	RX~J1940.1-1025	&	P	&	0.882	&	9.2	&	TRUE	\\
11	&	IGR~J15479-4529	&	IP	&	IGR~J15479-4529	&	IP	&	0.979	&	29.3	&	TRUE	\\
12	&	BY Cam	&	P	&	New	&	--	&	1.117	&	4.8	&	TRUE	\\
13	&	V2487 Oph	&	Nova	&	New	&	--	&	1.718	&	7.4	&	TRUE	\\
14	&	V1336 Aql	&	non-CV	&	4U~1916-053	&	LMXB	&	1.722	&	51.6	&	FALSE	\\
15	&	1RXS J0625+7334	&	IP	&	New	&	--	&	2.258	&	4.6	&	TRUE	\\
16	&	IGR~J00234+6141	&	CV	&	IGR~J00234+6141	&	CV	&	2.316	&	6.4	&	TRUE	\\
17	&	NGC 6440-CX8	&	CV?	&	4U~1745-203	&	LMXB	&	2.363	&	5.2	&	FALSE	\\
18	&	ISV 0115+63	&	Nova?	&	4U~0115+63	&	HMXB	&	2.502	&	348.9	&	FALSE	\\
19	&	GK Per	&	DN/IP	&	New	&	--	&	2.588	&	4.5	&	TRUE	\\
20	&	V2069 Cyg	&	CV 	&	New	&	--	&	2.808	&	5.3	&	TRUE	\\
21	&	V2400 Oph	&	IP	&	V2400 Oph	&	IP	&	3.120	&	22.9	&	TRUE	\\
22	&	V834 Cen	&	P	&	New	&	--	&	3.268	&	6.1	&	TRUE	\\
23	&	Sgr 	&	non-CV	&	Image structure	&	--	&	3.630	&	4.3	&	FALSE	\\
\hline
\end{tabular}
\end{minipage}
\end{table*}

\subsection[]{Correlation results}

Figure~\ref{correlation} plots the results of the correlations.  Also plotted are the mean of the number of matches obtained from the 4 fake catalogues and the difference between the matches obtained from correlating the Downes catalogue with the real IBIS excess list and the mean of the fake catalogue matches.  It can be shown that the optimum search radius is 4$\arcmin$, as after this point the fake matches increases in line with the matches obtained with the real IBIS data set. This corresponds well to the expected error on a faint IBIS detection \citep{Bird06}.  For a search radius of 4$\arcmin$ we obtain 23 matches (of which 4 are expected to be false); these are listed in Table~\ref{matches}. From inspection of the correlated sources,  match 23 is an image artefact in the IBIS maps and matches 1, 3, 4, 8, 14, 17 and 18 are positional coincidences between mainly globular cluster sources and previously identified X-ray objects, the majority of which are X-ray binaries presented in the second IBIS catalogue.  It is important to note that, working at the IBIS point source location accuracy of $\sim$2$\arcmin$, it is very difficult to associate a detection with an optical counterpart alone.   

Of the remaining 15 sources considered to be `true' matches, 9 are previously known as {\it INTEGRAL} detected CVs \citep[8 are included in the second IBIS catalogue: V709 Cas, IGR~J15479-4529, IGR~J17303-0601, V2400 Oph, V1223 Sgr,V1432 Aql/RX~J1940.1--102, IGR~J21335+5105, SS Cyg; IGR~J00234+6141 has been recently identified as a possible {\it INTEGRAL} discovered IP;][]{Masetti06b}.  The remaining 6 sources are new IBIS CV detections, which are GK Per, BY Cam, V834 Cen, V2487 Oph, V2069 Cyg and a new {\it INTEGRAL} source IGR~J06253+7334 whose {\it ROSAT} counterpart is an IP \citep[1RXS~J062518.2+733433;][]{Sofia03}.

\section[]{Properties of CVs detected by {\em INTEGRAL}}

Table~\ref{cvs} displays the characteristics of the CVs detected by {\it INTEGRAL}.  In addition to the 15 CVs identified through the correlation exercise, four IGRs have also been included as recently identified CV candidates \citep[IGR~J14536-5522, IGR~J15094-6649, IGR~J16167-4957, IGR~J17195-4100 from optical observations of the putative counterparts by][ and not yet included in the Downes catalogue]{Masetti06a}.  As expected, this sample of IBIS detected CVs contains mainly magnetic systems, 11 of which are either confirmed or probably IPs.  The remaining sources consist of 3 polars (2 of which are asynchronous), a single dwarf nova (SS Cyg) and 4 probable magnetic CVs. 

Many of the systems are very interesting detections.  V2400 Oph (also detected by {\it BeppoSAX}) is the sole example of a discless IP system \citep{Buckley95}.  A characteristic of discless IPs is that the accretion stream is expected to flip between the poles as the magnetic dipole rotates (Hellier 1991; Hellier \& Beardmore 2002).  V2487 Cyg has not been previously detected above 20~keV and was discovered as a fast nova in 1998.  This system was observed with {\it XMM-Newton} in 2001 by \citet{Hernanz02} and confirms the association with the {\it ROSAT} source RXS~J173200--191349 discovered in 1990.  This is a very interesting situation as it suggests the presence of X-ray emission from a CV before and after a nova erruption, with accretion being re-established as soon as $\sim$3 years after the nova event \citep{Hernanz02}. Although there has been no periodicity observed in the emission from V2487 Cen, the high-energy detection of this source by {\it INTEGRAL} provides further evidence that this is a intermediate polar system.

All but one of the systems with known periods have binary orbits greater than 3~hrs, placing them above the period gap.  Furthermore, 5 of these can be classed as long period systems with P$_{orb}$ greater than 7 hrs.  No significant modulations have been found in the 20--30~keV light curves, even when folded on known periods.  The majority of the CVs display faint persistent soft-gamma ray fluxes with the exception of V1223 Sgr and SS Cyg which are discussed later in this section.

In cases where no published distance was available, we obtained a rough
estimate by looking up the 2MASS K-band magnitudes and using an updated
version of Bailey's method \citep[][ Knigge, in prep.]{Bailey, Ramseyer}.
Our version of the method is calibrated against known CVs and implicitly
accounts for the expected contribution of the accretion luminosity to the
K-band flux. Distances obtained with this updated method are therefore not
strict lower limits but genuine estimates (albeit with relatively large
uncertainties). This method cannot be applied to long-period systems with
evolved donors ($>\sim$5~hrs) and therefore the estimates listed for V2069 Cyg and IGR J21335+5105 should be treated with caution.

\subsection[]{Average spectral properties}
In most cases, the CVs have unsufficient counts in an IBIS ScW to extract a spectrum using OSA.   In order to provide the most reliable estimates, survey-averaged source fluxes are obtained from the weighted mean of source light-curves generated from $>$11000 individual ScWs, as described in \citet{Bird06}.  By repeating this process for each source in 6 separate energy bands (20--30, 30--40, 40--60, 60--80 and 80--100~keV), time-averaged spectra for the observation period have been produced for each of the 19 sources listed in Table~\ref{cvs}.  The spectra have been fitted with both power law and thermal bremsstrahlung spectral models and the best-fit parameters are displayed in Table~\ref{fits} together with 20--100~keV flux and luminosity.  There are no significant differences between the best-fit spectral parameters of the CV sub-types detected by IBIS (Table~\ref{weight}).

%Furthermore, source fluxes (and hence spectra) cannot be reliably extracted directly from the IBIS survey mosaics as it has been estimated that a combination of the mosaicing process and the distortion of the sources in the mosaic can cause a reduction in the time-averaged source flux by $\sim$5\% \citep{Bird06}.

\begin{table*}
\centering
\begin{minipage}{140mm}
\caption{\label{cvs} {Table of characteristics of the CVs detected by IBIS}}
\begin{tabular}{@{}l l  c c c l l c }
\hline 
Name  & $\alpha$,$\delta$   & P$_{orb}$  & P$_{spin}$ & Dist\footnote{values in brackets have been estimated as described in the text}   & Type\footnote{IP=intermediate polar; P=polar; AP=asynchronous polar, DN=dwarf nova; N=nova; ?=unconfirmed classification} & X-ray & Refs \\
     & (IBIS position)    &       (min)     & (s)        & (pc)& & counterpart&\\
\hline

IGR~J00234+6141	&	5.688, 61.719			&	--	&	570?	&	300	&	IP?	&	1RXS~J002258.3+614111	&	1,2,3,4	\\
V709 Cas	&	7.252, 59.317			&	320.4	&	312.7	&	230$\pm$20	&	IP	&	RX~J0028.8+5917	&	5,6	\\
GK Per	&	52.820, 43.885			&	2875.2	&	351.34	&	420	&	DN/IP	& 		&	7,8	\\
BY Cam	&	85.743, 60.848			&	201.3	&	11846.4	&	190	&	AP	& 		&	7,9	\\
IGR~J06253+7334	&	96.317, 73.577			&	283	&	19.8	&	(500)	&	IP	&	1RXS~J062518.2+733433	&	10,11	\\
V834 Cen	&	212.262, -45.269			&	101.4	&	--	&	80	&	P  	&		&	7,12	\\
IGR~J14536-5522	&	223.406, -55.373			&	--	&	--	&	--	&	?	&	1RXS~J145341.1-5521	&	13,14,15	\\
IGR~J15094-6649	&	227.324, -66.849			&	--	&	--	&	--	&	?	&	1RXS~J150925.7-664913	&	15	\\
IGR~J15479-4529	&	237.033, -45.3484			&	562	&	693	&	$\sim$500	&	IP	&	1RXS~J15481.5--452845	&	16,17	\\
IGR~J16167-4957	&	244.155, -49.979			&	--	&	--	&	--	&	?	&	1RXS~J161637.2-495847	&	15	\\
IGR~J17195-4100	&	259.898, -41.051			&	--	&	--	&	--	&	?	&	1RXS~J171935.6--410054	&	15	\\
IGR~J17303-0601	&	262.593, -6.016		&	925.3	&	128	&	--	&	IP	&	1RXS~J173021.5--055933	&	18	\\
V2400 Oph	&	258.170, -24.267			&	205.2	&	927	&	(300)	&	IP	&	RX~J1712.6-2414	&	19	\\
V2487 Oph	&	262.968, -19.218			&	--	&	--	&	--	&	N/IP?	&	RXS~J173200--191349	&	20	\\
V1223 Sgr	&	283.755, -31.145			&	201.9	&	745.6	&	600	&	IP	&		&	7,21	\\
V1432 Aql	&	295.066, -10.408			&	201.94	&	12150.4	&	230	&	AP	&	RX~J1940.1--102 	&	22,23	\\
V2069 Cyg	&	320.899, 42.324			&	448.8	&	--	&	(1650)	&	IP?	&	RX~J2123.7+4217	&	24,25	\\
IGR~J21335+5105	&	323.375, 51.092			&	431.6	&	570.8	&	(1400)	&	IP	&	RX~J2133.7+5107	&	26	\\
SS Cyg	&	325.745, 43.587		&	396.2	&	--	&	166	&	DN	&		&	27,28	\\

\hline
\end{tabular}

\medskip
References: [1]~\citet{Masetti06b}; [2]~\citet{Halpern06}; [3]~\citet{Hartog06}; [4]~\citet{Bik06}; [5]~\citet{Bonnet01}; [6]~\citet{deMart01}; [7]~\citet{Ritter98}; [8]~\citet{Wu89}; [9]~\citet{Warner95}; [10]~\citet{Sofia03}; [11]~\citet{Staude03}; [12]~\citet{Berriman87}; [13]~\citet{Kuiper06}; [14]~\citet{Mukai06}; [15]~\citet{Masetti06a}; [16]~\citet{Haberl02}; [17]~\citet{deMart06}; [18]~\citet{Boris05}; [19]~\citet{Buckley95}; [20]~\citet{Hernanz02}; [21]~\citet{Bonnet82}; [22]~\citet{Watson95}; [23]~\citet{Geckeler97}; [24]~\citet{Motch96}; [25]~\citet{Thor01}; [26]~\citet{Bonnet06}; [27]~\citet{Friend90}; [28]~\citet{Harrison99}.

%\footnotetext[1]{Types without reference are from Downes et al. 2001.}\footnotetext[2]{Sekiguchi et al. 1992} $^{3}$Warner \& Woudt 2002;  $^{4}$Bonnet-Bidaud et al. 2001; $^{5}$Haberl et al.2002;  $^{6}$Buckley et al. 1995;  $^{7}$G{\"a}nsicke et al. 2005b; $^{8}$Ritter \& Kolb 1998; $^{9}$Watson et al.1995;  $^{10}$Bonnet-Bidaud et al. 2005;  $^{11}$Friend et al. 1990;  $^{12}$Thorstensen \& Taylor 2001;  $^{13}$de Martino et al. 2001; $^{14}$Geckeler \& Staubert 1997; $^{15}$Bonnet-Bidaud et al. 1982;  $^{16}$Harrison et al. 1999; $^{17}$Cropper 1990; $^{18}$Zuckerman et al.1992; $^{19}$Warner 1995;  $^{20}${Hernanz \& Sala 2002}
\end{minipage}
\end{table*}

\begin{table*}
%\footnotesize
%\renewcommand{\thefootnote}{\alph{footnote}}
 \centering
 \begin{minipage}{140mm}
  \caption{\label{fits}{20--100~keV best-fitted parameters for power law and thermal bremsstrahlung spectral models of the IBIS detected CVs ordered by sub-type. Luminosity in the 20--100~keV band has been calculated using observed 20--100~keV flux, distances in brackets represent luminosities calculated using uncertain distance estimates.}}
%\begin{center}
\begin{tabular}{@{}l c c c c c c c}
%\hline 
\hline  Name & $\Gamma$ & $\chi^{2}_{\nu}$ & kT  &  $\chi^{2}_{\nu}$ & Flux ($\times10^{-11}$ & Luminosity  & Type  \\
& &(3~d.o.f.) & (keV) & (3~d.o.f.) & erg~s$^{-1}$cm$^{-2}$)& ($\times10^{32}$erg~s$^{-1}$)&\\
\hline
%IGR~J00234+6141	&	3.1$\pm$0.5	&	5.7	&	15.9$\pm$5.1	&	6.1	&	1.1	&	1.18	&	IP?	\\
%V709 Cas	&	2.8$\pm$0.1	&	0.5	&	23.3$\pm$2.2	&	1.9	&	4.48	&	2.84	&	IP	\\
%GK Per	        &	2.7$\pm$0.7	&	3.5	&	28.7$\pm$15.6	&	3.3	&	2.54	&	5.36	&	DN/IP	\\
%BY Cam	        &	3.2$\pm$0.6	&	2.2	&	14.8$\pm$5.9	&	2.2	&	2.45	&	1.06	&	AP	\\
%IGR~J06253+7334	&	3.4$\pm$1.0	&	1.7	&	8.1$\pm$4.7	&	1.4	&	0.71	&	(2.12)	&	IP	\\
%V834 Cen	&	2.7$\pm$0.5	&	1.0	&	19.5$\pm$7.8	&	1.2	&	1.12	&	0.09	&	P  	\\
%IGR~J14536-5522	&	3.1$\pm$0.7	&	2.3	&	11.1$\pm$4.6	&	1.9	&	0.92	&	--	&	CV?	\\
%IGR~J15094-6649	&	3.6$\pm$0.8	&	1.0	&	13.8$\pm$5.1	&	1.0	&	1.38	&	--	&	CV?	\\
%IGR~J15479-4529	&	2.7$\pm$0.1	&	4.1	&	27.1$\pm$2.2	&	1.8	&	5.53	&	16.55	&	IP	\\
%IGR~J16167-4957	&	3.8$\pm$0.4	&	0.3	&	13.2$\pm$2.6	&	0.5	&	1.76	&	--	&	CV?	\\
%IGR~J17195-4100	&	2.9$\pm$0.2	&	0.9	&	27.0$\pm$4.4	&	1.2	&	2.46	&	--	&	CV?	\\
%IGR~J17303-0601	&	2.5$\pm$0.9	&	2.6	&	26.7$\pm$4.8	&	1.2	&	4.25	&	--      &	IP	\\
%V2400 Oph	&	3.1$\pm$0.1	&	4.5	&	18.6$\pm$1.4	&	5.0	&	3.29	&	(3.54)	&	IP 	\\
%V2487 Oph	&	3.0$\pm$0.5	&	2.2	&	25.5$\pm$8.6	&	1.9	&	1.03	&	--	&	Nova/IP?	\\
%V1223 Sgr	&	3.3$\pm$0.1	&	0.9	&	18.8$\pm$1.2	&	1.5	&	7.97	&	34.34	&	IP	\\
%V1432 Aql	&	2.8$\pm$0.1	&	0.2	&	25.4$\pm$7.0	&	0.5	&	3.37	&	2.13	&	AP	\\
%V2069 Cyg	&	2.4$\pm$0.4	&	1.5	&	35.7$\pm$16.8	&	1.7	&	1.35	&	(43.98)	&	IP?	\\
%IGR~J21335+5105	&	3.0$\pm$0.3	&	3.8	&	23.8$\pm$4.3    &	2.1	&	3.6	&	(84.44)	&	IP	\\
%SS Cyg	        &	3.2$\pm$0.2	&	2.9	&	13.6$\pm$8.1	&	3.4	&	2.78	&	0.92	&	DN	\\

SS Cyg	        &	3.2$\pm$0.2	&	2.9	&	13.6$\pm$8.1	&	3.4	&	2.78	&	0.92	&	DN	\\
V834 Cen	&	2.7$\pm$0.5	&	1.0	&	19.5$\pm$7.8	&	1.2	&	1.12	&	0.09	&	P  	\\
BY Cam	        &	3.2$\pm$0.6	&	2.2	&	14.8$\pm$5.9	&	2.2	&	2.45	&	1.06	&	AP	\\
V1432 Aql	&	2.8$\pm$0.1	&	0.2	&	25.4$\pm$7.0	&	0.5	&	3.37	&	2.13	&	AP	\\
V709 Cas	&	2.8$\pm$0.1	&	0.5	&	23.3$\pm$2.2	&	1.9	&	4.48	&	2.84	&	IP	\\
GK Per	        &	2.7$\pm$0.7	&	3.5	&	28.7$\pm$15.6	&	3.3	&	2.54	&	5.36	&	DN/IP	\\
IGR~J06253+7334	&	3.4$\pm$1.0	&	1.7	&	8.1$\pm$4.7	&	1.4	&	0.71	&	(2.12)	&	IP	\\
IGR~J15479-4529	&	2.7$\pm$0.1	&	4.1	&	27.1$\pm$2.2	&	1.8	&	5.53	&	16.55	&	IP	\\
IGR~J17303-0601	&	2.5$\pm$0.9	&	2.6	&	26.7$\pm$4.8	&	1.2	&	4.25	&	--      &	IP	\\
V2400 Oph	&	3.1$\pm$0.1	&	4.5	&	18.6$\pm$1.4	&	5.0	&	3.29	&	(3.54)	&	IP 	\\
V1223 Sgr	&	3.3$\pm$0.1	&	0.9	&	18.8$\pm$1.2	&	1.5	&	7.97	&	34.34	&	IP	\\
IGR~J21335+5105	&	3.0$\pm$0.3	&	3.8	&	23.8$\pm$4.3    &	2.1	&	3.6	&	(84.44)	&	IP	\\
IGR~J00234+6141	&	3.1$\pm$0.5	&	5.7	&	15.9$\pm$5.1	&	6.1	&	1.1	&	1.18	&	IP?	\\
V2069 Cyg	&	2.4$\pm$0.4	&	1.5	&	35.7$\pm$16.8	&	1.7	&	1.35	&	(43.98)	&	IP?	\\
V2487 Oph	&	3.0$\pm$0.5	&	2.2	&	25.5$\pm$8.6	&	1.9	&	1.03	&	--	&	Nova/IP?	\\
IGR~J14536-5522	&	3.1$\pm$0.7	&	2.3	&	11.1$\pm$4.6	&	1.9	&	0.92	&	--	&	?	\\
IGR~J15094-6649	&	3.6$\pm$0.8	&	1.0	&	13.8$\pm$5.1	&	1.0	&	1.38	&	--	&	?	\\
IGR~J16167-4957	&	3.8$\pm$0.4	&	0.3	&	13.2$\pm$2.6	&	0.5	&	1.76	&	--	&	?	\\
IGR~J17195-4100	&	2.9$\pm$0.2	&	0.9	&	27.0$\pm$4.4	&	1.2	&	2.46	&	--	&	?	\\

\hline

\end{tabular}
\end{minipage}
\end{table*}

%\begin{figure}\label{}
%\includegraphics[width=84mm]{}
% \caption{}
%\end{figure}

%\begin{figure*}
%\vbox to 220mm{\vfil Landscape figure to go here. This figure was
%not part of the original paper and is inserted here for
%illustrative purposes.\\ See the author guide for details (section
%2.2 of \verb|mn2eguide.tex|) on how to handle landscape figures or
%tables. \caption{} \vfil} \label{landfig}
%\end{figure*}

\begin{table}
%\footnotesize
%\renewcommand{\thefootnote}{\alph{footnote}}
%\begin{center}
\caption{\label{weight} {Weighted means of 20--100~keV spectral fits to power law ($\Gamma$) and thermal bremsstrahlung (kT) models of IBIS detected CVs by sub-type}}
\begin{tabular}{@{}l c l  }
\hline 
Type & $\Gamma$ & kT (keV) \\
\hline
IP & 2.9$\pm$0.1 & 20.9$\pm$0.8 \\
P & 2.8$\pm$0.2 & 19.3$\pm$3.9 \\
DN (SS Cyg) & 3.2$\pm$0.242 & 13.6$\pm$8.1 \\
\hline
\end{tabular}
%\renewcommand{\baselinestretch}{1.0}\normalsize
%\end{center}
\end{table}

\subsection[]{V1223 Sgr}

The IP system V1223 Sgr is the most significantly detected CV in the current IBIS survey maps, with a significance of 38$\sigma$ in the 20--40~keV final mosaic.  This system has also been detected by {\it RXTE} between 1996 and 2000 and the broad-band spectrum ({\it RXTE}/HEXTE + {\it INTEGRAL}/IBIS, 3--100~keV) provides an estimate of the post-shock temperature of kT=29$\pm$2~keV when fit with a thermal bremsstrahlung model with reflection from an optically thick cold medium plus neutral absorption (n{$_{\textrm{H}}$=3.3$\times$10$^{22}$~cm$^{-2}$) and an iron line at $\sim$6.5~keV \citep{Revnivtsev04}. 

\begin{figure}
\includegraphics[totalheight=3.0cm]{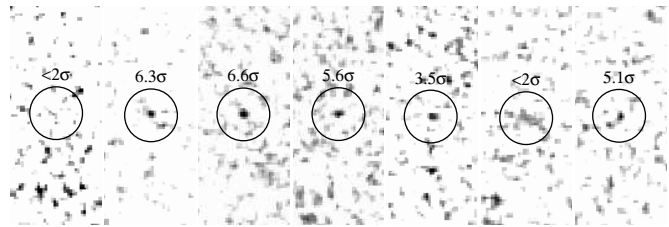}
\caption{\label{V1223image} {20--30~keV images of IBIS revolution 61, ScWs 97--103 (left to right).  The location of V1223 Sgr is circled and the significances labelled for each ScW.}}
\end{figure}
\begin{figure}
\includegraphics[totalheight=8.5cm, angle=-90]{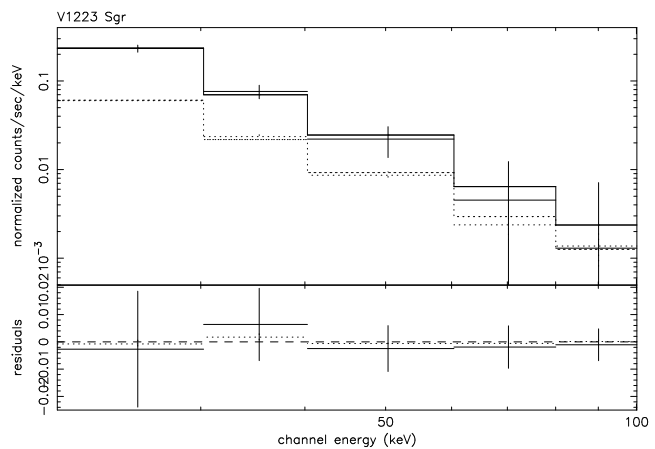}
\caption{\label{V1223spec} {20--100~keV spectra and best-fitted thermal bremsstrahlung models of V1223 Sgr for the burst (solid) and survey average (dotted).}}
\end{figure}
Interestingly, inspection of the 20--30~keV IBIS survey lightcurve yields an episode of intense brightening of the source lasting for $\sim$3.5~hrs during revolution 61 (MJD=52743).  The source is clearly visible in a single ScW with a significance of up to 6.5$\sigma$ and peak flux approximately three times that of the average (see Fig.~\ref{V1223image}). It is notable that the burst timescale is very similar to the orbit of the binary (P$_{orb}$=3.36~hr).  A burst spectrum has been extracted from ScW 98 and 99 in revolution 61 and is compared to the average survey spectrum in Fig.~\ref{V1223spec}. The best-fitting bremsstrahlung temperature of the burst in the 20--100~keV energy band is slightly softer compared to the average spectrum (13.78$\pm$2.50~keV and 18.77$\pm$1.20~keV, respectively).  Both temperatures are significantly cooler than the temperature estimated by \citet{Revnivtsev04},  kT=29~keV, derived from 3--100~keV spectrum.   The 20--100~keV luminosity has been calculated and is seen to increase from 3.43$\times$10$^{33}$~erg~s$^{-1}$ as measured from the average survey flux, to 1.13$\times10^{34}$~erg~s$^{-1}$ over the 3.5~hr burst period.
%A sub-ScW light-curve has been produced for the 7 ScW in which the burst occured, shown in Fig.~\ref{V1223lc}

Such behaviour is previously unreported in the high-energy regime for mCVs. However, an unusual short-term burst has also been detected from this system in the optical by \citet{Amerongen89}.  The single optical outburst lasted between 6 and 24~hrs and resulted in an increase in flux by a factor of 3.  Van Amerongen \& van Paradijs suggest that the outburst could be comparable to a DN outburst, but made shorter by the absence of the inner part of the accretion disk as seen in moderately magnetic IPs.  Similar short optical oubursts have also been observed from another IP, TV Col \citep{Angelini89,  Hellier93}. It is not clear whether these outbursts are the result of disk instabilities or an increase in mass transfer from the secondary.  If the high energy emission is assumed to originate from the hot, shocked plasma in the accretion column, the outburst observed from V1223 Sgr must be a result of a temporary increase in accretion rate onto the WD.  The Optical Monitoring Camera (OMC) on-board {\it INTEGRAL} detected two fast flares from V1223 Sgr ($\sim$15~min and $\sim$2.5~hr) one year later (MJD=53110; 53116) than the IBIS burst \citep{Simon05}.  Contemporaneous IBIS observations yield no increase in flux at ScW level at these times.  No OMC detected flares are reported during the epoch of the IBIS burst and no significant increase in X-ray emission is seen during the IBIS burst period, as detected by the All-sky Monitor (ASM) on {\em{RXTE}} (quick-look results provided by the ASM/{\em{RXTE}} team, 1.5--12~keV).  \citet{Beardmore00} present evidence for complex intrinsic absorption in X-ray observations of V1223 Sgr. Continual monitoring of this source is important to identify whether this activity is recurrent.

\subsection[]{SS Cyg}
SS Cyg is an optically bright DN observed to undergo outbursts every $\approx$40 days, characterised by an increase in optical magnitude from V$\sim$12 to V$\sim$8.  The X-ray generation mechanism in DNe predicts a suppression of high energy emission during the regular optical outburst phases that typify these objects.  The 20-30~keV IBIS light-curve has been correlated with the optical light-curve (validated AAVSO\footnotemark[1] \footnotetext[1]{\texttt{http://www.aavso.org/}}data) to determine if the soft gamma-ray emission changes during the DN outbursts.  Fig.~\ref{SSlc} plots the optical magnitude and 20--30~keV count rate over part of the IBIS survey period and suggests a strengthening of the 20--30~keV flux during optical quiesence.  This confirms the finding of \citet{Mcgowan04} from correlated X-ray ({\it RXTE}}/ASM; 3--12~keV) and optical emission.  

IBIS spectra have been extracted during optical outburst (MJD~= 53036.2--53041.5) and quiescence (MJD = 53018.7--53023.8) to quantify any spectral variation and are shown in Fig.~\ref{SSspec} with the best fit bremsstrahlung model.  The source is very faint during optical outburst and the data cannot be fitted satisfactorily by either this model or a simple power law.  The best-fitting thermal bremmstrahlung temperature (kT = 17.43~$\pm$2.7~keV) and 20~keV flux levels detected by IBIS during optical quiescence are consistent with previously published 3--20~keV {\it {RXTE}}/PCA values \citep{Mcgowan04}.  The 20--100~keV flux is of an order of magnitude greater during optical quiescence than outburst as detected by IBIS;  L(20-100~keV) increases from 1.4$\times$10$^{31}$~erg~s$^{-1}$ to 2.3$\times10^{32}$~erg~s$^{-1}$.  This result confirms previous findings whereby hard X-ray flux is suppressed during optical outbursts \citep{Wheatley03, Mcgowan04} and also provides evidence for this behaviour extending to emission beyond 20~keV. Further work with an increased IBIS data set should be able to provide a more detailed cross-correlation between other wavebands and help track the time-lags between the optical, X-ray and soft gamma-ray emission.

%it is interesting to see that the suppression of hard X-rays extends up to 100~keV for this system, suggesting that the origin of the 20--100~keV emission is indeed from the same region/same emission mechanism as the $<$20~keV X-ray emission.  

\begin{figure}
\includegraphics[totalheight=8.5cm, angle=-90]{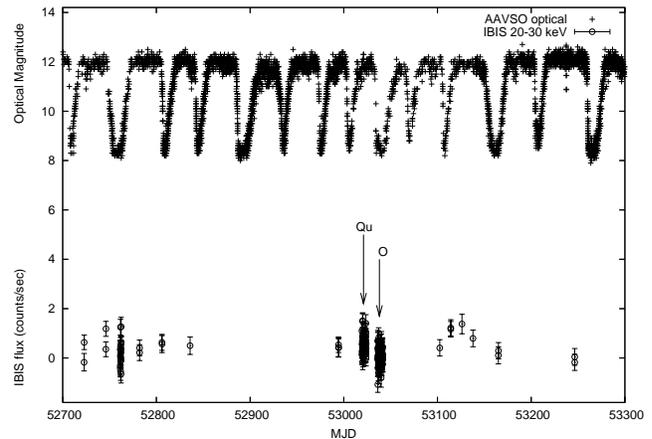}
\caption{\label{SSlc} {SS Cyg light curve using optical data from AAVSO (crosses) and 20--30~keV IBIS survey data (circles). The arrows point to groups of IBIS data that are consistent with periods of optical outburst (O) and quiescence (Qu) from which spectra have been extracted.}}
\end{figure}

\begin{figure}
\includegraphics[totalheight=8.5cm, angle=-90]{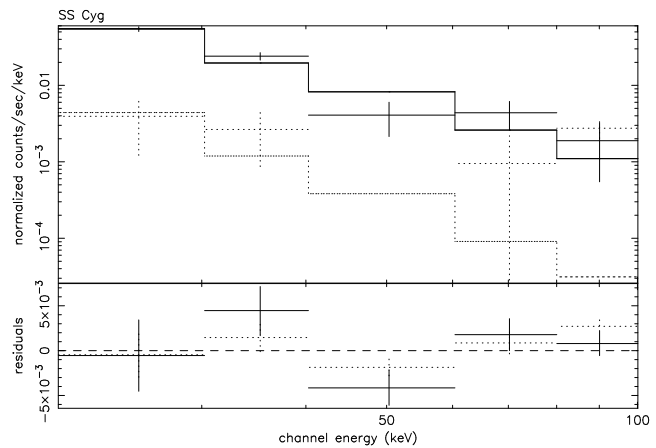}
\caption{\label{SSspec} {20-100~keV spectra and best-fitted thermal bremsstrahlung models of SS Cyg during optical outburst (dotted) and optical quiescence (solid).}}
\end{figure}

\section[]{Discussion}

\begin{figure*}
\centering
 \begin{minipage}{140mm}
%\begin{center}
\includegraphics[totalheight=9cm]{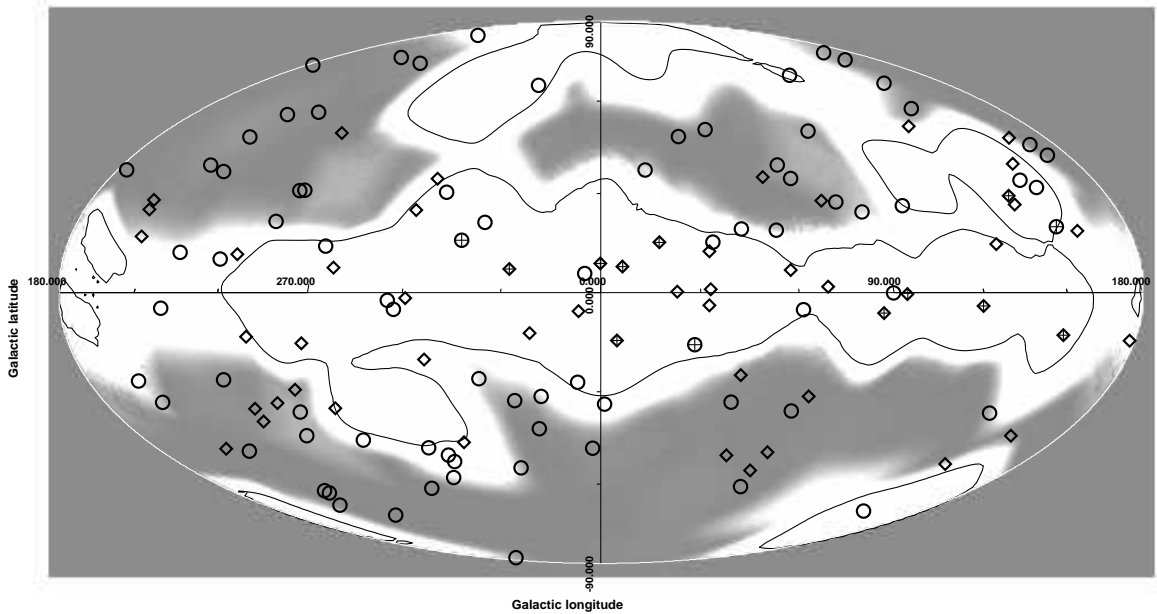}
\caption{\label{exp} {Catalogued magnetic CVs overlaid on the IBIS survey exposure map.  Contour represents 150~ksec of exposure.  Diamonds are intermediate polars; circles are polars; sources additionally filled with crosses signify IBIS detections.}}
%\end{center}while for asynchronous polars this is
\end{minipage}
\end{figure*}

%Swift~J0732.5--1331, not seen by IBIS.  Swift also sees DO Dra and PQ Gem (both IPs).  IGR J14536-5522 identified also by Swift and from X-ray spectra could be mCV?

%Blurring between boundaries of different magnetic CV classes

IBIS has so far detected a group of 19 CVs, the majority of which are intermediate polars, but also including a bright DN, a rare discless IP and 2 asychronous polars.  Despite the range of types, this paper has shown that the spectral characteristics of these objects in the 20--100~keV band are actually quite similar and compare well with previous high-energy spectral fits and luminosity estimates of magnetic CVs \citep{deMart04, Suleimanov05}. The spectra of most of the detected objects can be equally well fit with a power law or a thermal bremsstrahlung model ($\Gamma\sim$2.8; kT$\sim$20~keV) which also provides an estimate of the temperature of the post shock region.  As a result of the uncertainty of many of the source distance estimates, a detailed analysis of the relative 20--100~keV luminosities of the source types is not possible, but from the values listed in Table 3, it can be shown that generally IPs are more luminous than polars (IPs:~L$\approx>$2$\times$10$^{32}$~erg~s$^{-1}$; polars:~L$\approx<$2$\times$10$^{32}$~erg~s$^{-1}$) and in some cases are an order of magnitude more luminous. Furthermore, the asynchronous polars are an order of magnitude more luminous than the single sychronous polar detected by IBIS (V834 Cen, L$\sim$10$^{31}$~erg~s$^{-1}$).

%(although the polars appear fainter than IPs (Mean F$_{20-100keV}\sim$2.3$\times10^{-11}$ and 4.8$\times10^{-11}$~erg~s$^{-1}$cm$^{-2}$ respectively).) 
%Table~\ref{weight} lists the 20--100~keV luminosity for a selection of sources.

%\begin{landscape}
%\begin{figure}
%\begin{center}
%\includegraphics[totalheight=2.6cm]{exp2.ps}
%\renewcommand{\baselinestretch}{1.0}\normalsize
%\caption{\label{exp2} {Polars (blue) and IPs (green) detected within 250~ksec of IBIS exposure (contour).  IBIS detected sources (polars or IPs) are marked with a red cross.}}
%\end{center}
%\end{figure}
%\end{landscape}

The preferential detection of IPs by IBIS is not unexpected.  As suggested by \citet{Kuijpers82} the reason for the lower levels of hard X-ray/soft gamma-ray emission from polars may be related to the low \.M$_{acc}$ and the high magnetic field.  It is tempting therefore to conclude that the polars detected by IBIS, by virtue of the existence of $>$20~keV emission, have a lower B field and/or higher rate of accretion (therefore creating a less `blobby' flow) than compared with non detected polars.  Moreover, two of the three polars observed by IBIS do not rotate sychronously.  Asynchronous polars are a very rare sub-class of CVs consisting of just 4 confirmed members.  The remaining 2 asynchronous polars are not currently detected by IBIS, but are located in regions of relatively low exposure.  Asynchronicity in polars has been interpreted as a disruption due to a nova eruption \citep{Schmidt88, Mukai03} or spin up of the WD caused by accreted angular momentum in high \.M$_{acc}$ systems \citep[e.g. {\it INTEGRAL}/IBIS detected polar, BY Cam,][]{Silber92, Schwarz05}.  From the study of Doppler tomograms, \citet{Schwarz05} have shown that the optical emission of BY Cam probably originates from an accretion curtain, as opposed to a focused accretion stream.  They suggest that this implies that the accreted material enters the magnetosphere of the WD without being coupled to the magnetic field and predict an accretion rate a factor of two higher than expected from the orbital period of this system ($\sim$3.3~hr).   According to the standard theory of CV evolution, CVs with P$_{orb}<$3~hr are older and have lower \.M$_{trans}$ than longer P$_{orb}$ systems as a consequence of disrupted magnetic braking \citep{Kolb96, Boris05}. Interestingly, the least luminous of the polars detected by IBIS, V834 Cen, is a more typical phase-locked, short P$_{orb}$ system. Apart from this source, all the other {\it INTEGRAL} CVs have P$_{orb}>$3~hr.  The unusual periodicity of the asynchronous polar V1432 Aql/RX~J1940.1--1025 (P$_{spin}>$P$_{orb}$), also observed by IBIS, has given rise to suggestions that it should be reclassified as an IP \citep{Rana05}.  The argument is supplemented by the strong hard X-ray emission detected from this source. 

It is useful to look at the distribution of the CVs detected by {\it INTEGRAL} compared to the complete set of catalogued magnetic CVs.  Figure~\ref{exp} is an exposure map of the IBIS survey data set used in this paper.  All IBIS CVs detected above 4$\sigma$ in the 20--100~keV energy range are found in regions for which the exposure is 150~ksec or above (contour marked), the vast majority in areas of at least 250~ksec.  Overlaid on the exposure map are the locations of all currently known magnetic CVs (including those with unconfirmed polar or IP classifications), as currently listed on the online CV database CVcat\footnotemark[1]\footnotetext[1]{{\texttt{http://www.cvcat.net/}}}.  Sources additionally marked with a cross are the magnetic CVs detected by IBIS.  A total of 25 IPs (of which 12 are unconfirmed) and 17 polars (of which 2 are unconfirmed) are located within the 150~ksec contour.  From this simple analysis it can be seen that a significant proportion ($>$60~per cent) of the confirmed IPs have been detected by IBIS in the 20--100~keV band.  Of the 16 catalogued IPs not detected by IBIS, 12 are listed as unconfirmed IP classifications.  The picture is somewhat different for polars:  IBIS detects only 3, leaving 11 confirmed polars undetected, a detection rate of $\sim$20~per cent.  This appears to confirm the observational evidence that IPs are harder sources than polars, as IBIS seems to preferentially detect IPs (all newly identified CVs by {\it INTEGRAL} are IPs) and asynchronous polars.  Consequently, deeper observations and exposure away from the Galactic plane should reveal more IP detections by IBIS.  

\citet{Boris05} has suggested that {\it INTEGRAL} might be a useful tool for detecting new IPs, complementing the (mainly polar) {\it ROSAT} discoveries in the X-ray band ($<$10~keV).  The present study confirms this.  Particularly in the Galactic plane and centre, absorption inhibits the detection of these source from previous X-ray surveys, making IBIS an ideal resource for targeting the relatively X-ray hard IPs.  Optical follow-up is, however, imperative to understand whether some of the wealth of unidentified sources are, in fact, CVs.  It is possible that a proportion of the $\sim$50 unclassified {\it INTEGRAL} sources are unidentified CVs. 

\section[]{Conclusions}
This paper has presented the detection of 19 CVs by the imager on board {\it INTEGRAL} in the 20--100~keV energy range by mosaicing almost three years of survey data.  Of these objects, 9 are previously known {\it INTEGRAL} CV detections (8 from IBIS cat 2 + IGR~J00234+6141) and a further 6 objects have been discovered in IBIS survey mosaics, including a new {\it INTEGRAL} source IGR~J06253+7334.  An additional 4 IGR sources are included in the analysis as recently identified CV candidates (IGR~J14536-5522, IGR~J15094-6649, IGR~J16167-4957, IGR~J17195-4100) which are expected to be new IPs \citep{Masetti06a}.  As with previous high-energy detections of CVs, it is found that the source sample consists mainly of magnetic systems (8 IPs, 3 unconfirmed IPs and 3 polars).  Interestingly, 2 of the three polar systems are classed as asynchronous and one of the IPs is discless; both are extremely rare types of objects.  Analysis of long term light-curves generally yields faint, persistent soft gamma-ray fluxes, with two exceptions.  Firstly, evidence is obtained of an unusual short-timescale ($\sim$4~hrs) burst from an IP, V1223 Sgr, the first such detection from this type of object in the high-energy regime.  The 20--100~keV lumininosity is seen to increase to $\sim$10$^{34}$~erg~s$^{-1}$ compared to an average level of $\sim$3$\times$10$^{33}$~erg~s$^{-1}$. Secondly, contemporaneous optical and soft gamma-ray observations of the bright dwarf nova SS Cyg show a significant decrease in 20-100~keV flux/luminosity during optical outburst.  Similar results have been obtained from the correlation of X-ray and optical observations of SS Cyg which support the theory that hard X-ray emission is suppresed during periods of high mass accretion \citep{Pringle79, Wheatley03}. 

The general spectral characteristics of all the IBIS CVs are presented and it is shown that the 20--100~keV spectra can be fit equally well with a power law or thermal bremsstrahlung model with no significant difference between the average best-fitting parameters of IP and polar spectra.  Average 20--100~keV luminosity estimates of IPs are an order of magnitude higher than for polars, but this is dependent upon very limited distance information.   Finally, analysis of the location of these objects compared to the exposure bias of the IBIS survey campaign leads to the conclusion that IBIS is especially suitable for the detection of IPs and asynchronous polars.  As a group, IPs constitute only about 2\% of the total number of catalogued CVs, but from the results of recent observations targetting candidate {\it ROSAT} and {\it INTEGRAL} CVs, it is very likely that this number will grow.  {\it INTEGRAL} is enabling additional {\it ROSAT} sources to be targeted for IP identification (all IGR CVs have {\it ROSAT} counterparts).  It can be expected that with more comprehensive sky coverage and increased exposure, the number of CVs (most likely more IPs) detected by IBIS will increase further and that continuing optical and X-ray follow-ups of unclassified {\it INTEGRAL} sources will lead to the discovery of new IPs.

\section*{Acknowledgments}
This work is based on observations with {\it INTEGRAL\/}, an ESA project with instruments and science data centre funded by ESA member states (especially the PI countries: Denmark, France, Germany, Italy, Switzerland, Spain), Czech Republic and Poland and with the participation of Russia and the USA.  Funding for the UK collaborators is provided by PPARC. The authors would like to thank N. Masetti for providing information about optical observations of {\it INTEGRAL} sources, K.E. McGowan for useful discussions.  We would also like to thank the referee for their prompt and helpful comments.

\bsp

\label{lastpage}

\end{document}